\documentclass{article}

\usepackage{graphicx}
\usepackage{amsmath}
\usepackage{amsfonts}
\usepackage{amssymb}
\usepackage{amsthm}
\usepackage{booktabs}
\usepackage{float}
\usepackage[round]{natbib}
\usepackage[margin=1in]{geometry}
\usepackage{authblk}
\usepackage{colortbl}
\usepackage[dvipsnames]{xcolor}
\usepackage[
  colorlinks=true,
  linkcolor=blue,
  citecolor=blue!60!black,
  urlcolor=blue!60!black,
  filecolor=black
]{hyperref}

\usepackage{tikz}
\usetikzlibrary{positioning, decorations.pathreplacing}

\newcommand{\E}{\mathbb{E}}

\newcommand{\TTE}{\text{TTE}}

\title{
Validating Causal Message Passing Against Network-Aware Methods on Real Experiments}

\author{Albert Tan{\textsuperscript{*}}\quad
Sadegh Shirani{\textsuperscript{$\dagger$}}\quad
James Nordlund{\textsuperscript{*}}\quad
Mohsen Bayati{\textsuperscript{*,$\dagger$}}
}
\date{}

\begin{document}
\maketitle
\begingroup
  \renewcommand\thefootnote{}
  \footnotetext{\textsuperscript{*}\;Amazon \quad \textsuperscript{$\dagger$}\;Stanford University}
  \addtocounter{footnote}{-1}
\endgroup

\begin{abstract}
Estimating total treatment effects in the presence of network interference typically requires knowledge of the underlying interaction structure. However, in many practical settings, network data is either unavailable, incomplete, or measured with substantial error. We demonstrate that causal message passing, a methodology that leverages temporal structure in outcome data rather than network topology, can recover total treatment effects comparable to network-aware approaches. We apply causal message passing to two large-scale field experiments where a recently developed bipartite graph methodology, which requires network knowledge, serves as a benchmark. Despite having no access to the interaction network, causal message passing produces effect estimates that match the network-aware approach in direction across all metrics and in statistical significance for the primary decision metric. 
Our findings validate the premise of causal message passing: that temporal variation in outcomes can serve as an effective substitute for network observation when estimating spillover effects. This has important practical implications: practitioners facing settings where network data is costly to collect, proprietary, or unreliable can instead exploit the temporal dynamics of their experimental data.
\end{abstract}

\section{Introduction}\label{sec:intro}

Randomized experiments form the foundation of evidence-based decision-making across domains ranging from technology services to public health programs. The classical analysis of such experiments relies on the Stable Unit Treatment Value Assumption (SUTVA), which posits that each unit's outcome depends only on its own treatment assignment \citep{cox1958planning}. However, in networked settings, where units interact through social ties, shared resources, or competitive dynamics, this assumption fails. A unit's outcome may depend not only on its own treatment but also on the treatments received by connected units, a phenomenon known as \emph{network interference} or \emph{spillover} \citep{halloran1995causal, sobel2006what, hudgens2008toward}.

The methodological response to interference has followed two broad paths. The first path leverages explicit knowledge of the network structure, developing estimators that condition on or adjust for network exposure \citep{aronow2017general, leung2022causal}. Recent advances in bipartite experimental designs exemplify this approach: when treatment units (e.g., service providers) and outcome units (e.g., customers) form a two-sided structure, methods can exploit the observed interaction graph to define exposure mappings and construct interference-aware estimators \citep{pouget2019variance,doudchenko2020causal,zigler2021bipartite,harshaw2023design,tan2025estimating}. These network-aware methods achieve strong performance when the interaction structure is observed, but they require data that is often unavailable, proprietary, or measured with substantial error.

The second path addresses settings where the network is unknown or only partially observed. Recent work has developed methods that substitute temporal or design-based variation for network knowledge \citep{johari2022experimental, cortez2022staggered,munro2021treatment}. Among these, \emph{causal message passing} (CMP) offers a principled framework that leverages a key insight: the mechanisms governing how treatment effects propagate through a network exhibit structural invariance over time \citep{shirani2024causal, shirani2025can, shirani2025evolution}. By observing how outcomes evolve across time periods, CMP learns the aggregate dynamics of interference without requiring observation of individual network connections.

This paper bridges these two methodological streams by empirically validating CMP against a network-aware benchmark. We apply CMP to two large-scale field experiments previously analyzed using the bipartite graph methodology of \citet{tan2025estimating}, which explicitly incorporates the interaction structure between treatment and outcome units. Our main finding is that CMP, despite operating without any network information, produces total treatment effect estimates that closely match those from the network-aware approach and correctly recover the direction of interference bias predicted by domain knowledge.

\begin{figure}[t]
\centering
\includegraphics[width=\textwidth]{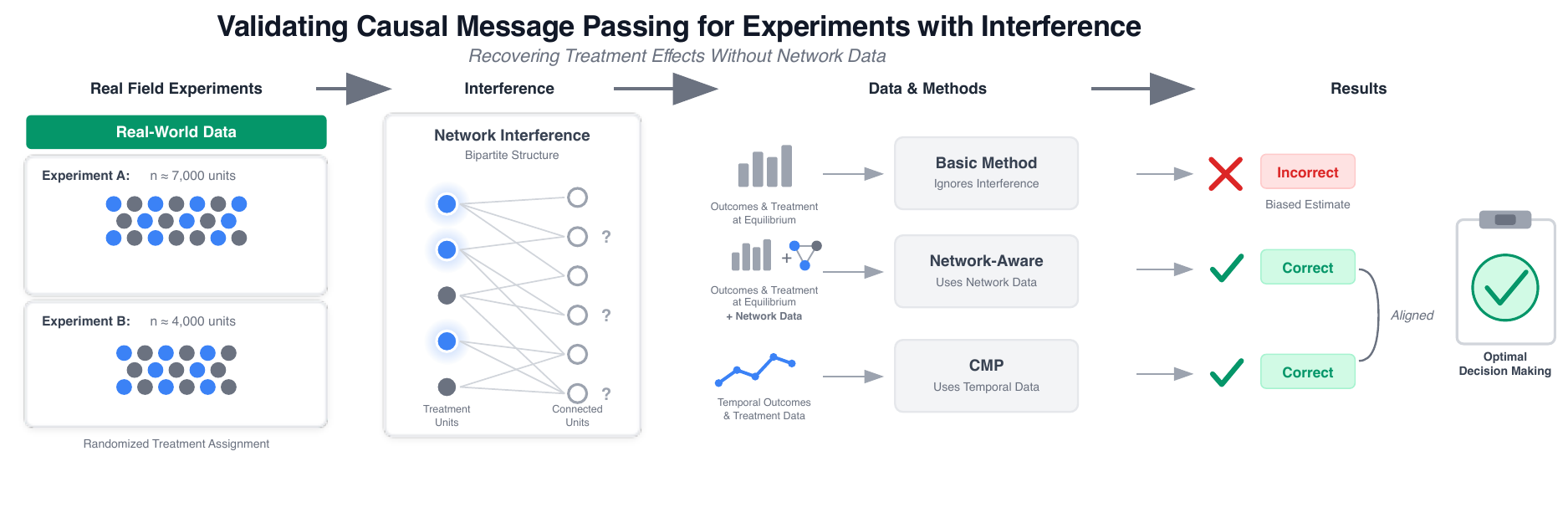}
\caption{Overview of our empirical validation framework. We apply three estimation methods to two real field experiments (Experiment A: $n \approx 7{,}000$; Experiment B: $n \approx 4{,}000$) conducted in a bipartite setting where treatment units interact with connected units, creating network interference. The \emph{Basic Method} ignores interference and produces biased estimates. The \emph{Network-Aware} approach leverages observed network data to correct for spillovers. \emph{Causal Message Passing (CMP)} uses only temporal outcome dynamics, without any network information, yet produces estimates that align with Network-Aware in both direction and significance on the primary decision metric.}
\label{fig:overview}
\end{figure}

\subsection{Contributions}

Our contributions are:

\begin{enumerate}
    \item \emph{Empirical validation of CMP.} We demonstrate that CMP recovers total treatment effects comparable to network-aware methods on real experimental data. This provides the first direct comparison of these methodological approaches on real world field experiments.
    
    \item \emph{Bias correction without network data.} We show that CMP correctly identifies the direction of interference bias for metrics where economic theory provides directional predictions. Like the network-aware benchmark, CMP reverses the misleading conclusions that would arise from naive difference-in-means estimation.

\item \emph{Practical guidance.} Our results suggest that practitioners facing settings where network data is costly, proprietary, or unreliable can employ CMP as a viable alternative. The temporal dynamics of experimental outcomes contain sufficient information to adjust for interference effects.
\end{enumerate}

\subsection{Related Work}

Understanding and estimating the effects of treatments and interventions is a foundational problem across many fields, and has attracted increasing attention in recent years \citep{halloran1995causal,sobel2006what,hudgens2008toward,tchetgen2012interference,manski2013identification,aronow2017general,saveski2017detecting,athey2018exact,viviano2020experimental,hu2022average,han2022detecting,leung2022causal,agarwal2022network,belloni2022neighborhood,li2022network,li2022random,farias2022markovian,candogan2023correlated,ni2023design,viviano2023causal,harshaw2023design,imbens2024causal,eichhorn2024low,johari2024does,Larsen02042024,peng2025differencesinneighborsnetworkinterferenceexperiments}.

The current work connects several strands of the interference literature, ranging from network-aware designs to settings with unknown network structure and to marketplace experiments. Specifically, network-aware experimental designs have been developed by \citet{ugander2013graph} and \citet{eckles2017design}, while bipartite interference settings have received substantial attention in \citep{pouget2019variance,doudchenko2020causal,zigler2021bipartite,NEURIPS2022_f7f043c3,harshaw2023design,tan2025estimating}.

In the context of service experiments, a growing body of research has examined interference effects across a range of economic environments and experimental designs \citep{blake2014marketplace,holtz2020reducing,wager2021experimenting,munro2021treatment,johari2022experimental,10.1145/3485447.3512063,bajari2023experimental,farias2023correcting,zhu2024seller,wu2024switchbackpriceexperimentsforwardlooking,bright2022reducing,zhang2025debiasing,johari2025estimationtreatmenteffectsnonstationarity}. Leveraging temporal observations and structure has also proven promising for addressing interference \citep{farias2022markovian,hu2022switchback,bojinov2023design,li2023experimenting,ni2023design,han2024population,jia2025clustered}.

Methods for settings with unknown network structure include mean-field-type approaches \citep{wager2021experimenting,johari2022experimental,munro2021treatment}, design-based experimental strategies \citep{cortez2022staggered,hu2022switchback}, and total-effect estimators that do not require explicit network information \citep{yu2022estimating}. This paper utilizes a recent methodology in this stream, particularly the causal message passing (CMP) framework introduced in \citet{shirani2024causal} and extended to settings with higher-order dynamics in \citet{bayati2024higher}. While these initial versions of CMP perform well when sufficient temporal or cross-sectional variation is available, their effectiveness can be limited in short experiments or in environments with complex outcome-treatment dynamics. To address this challenge, \citet{shirani2025can} proposes a \emph{distribution-preserving network bootstrap} that constructs multiple representative subpopulations from a single experiment, thereby increasing effective sample size for estimation; we adopt this approach in the present work. More recently, \citet{shirani2025evolution} provides an axiomatic foundation for evolution-based methods through an exposure-mapping perspective.
Our contribution is to provide a direct empirical comparison between bipartite, network-aware designs and network-blind causal message passing using real experimental data.

\section{Methodology Overview}\label{sec:methods}

This section provides summaries of the two methodologies we compare: the network-aware bipartite approach and causal message passing. But we defer for a detailed description to the original papers. Specifically, to \citep{tan2025estimating} and \citep{shirani2024causal,shirani2025can}. Both target the same estimand, the total treatment effect under full deployment, but differ fundamentally in their information requirements.

\subsection{Estimand: Total Treatment Effect}

Consider an experiment with $N$ units indexed by $i = 1, \ldots, N$, observed over time periods $t = 0, 1, \ldots, T$. Let $W_t^i \in \{0,1\}$ denote the treatment assignment for unit $i$ at time $t$, and let $Y_t^i(\mathbf{W})$ denote the potential outcome for unit $i$ at time $t$ under the full treatment allocation matrix $\mathbf{W}$, where $\mathbf{W}$ is an $N \times T$ matrix with entry $(i,t)$ being $W_t^i$.

When running an experiment, practitioners face choices about how treatment assignments evolve over time. One approach fixes each unit's treatment status throughout the experiment: units are assigned to treatment or control at the outset, and experimenters observe outcomes as they stabilize to an equilibrium reflecting the treatment's full effect. An alternative approach allows treatment assignments to vary over time, with units potentially switching between treatment and control across periods. A particularly common design in this latter category is the \emph{staggered rollout}, where all units begin in the control state, and at some point during the experiment, units transition to treatment and remain treated thereafter. In this paper, we primarily focus on settings consistent with staggered rollout designs. Consequently, both outcomes and treatments are naturally represented as $N \times T$ matrices, capturing the temporal evolution of each unit's status and response.

The \emph{total treatment effect} (TTE) compares the average outcome under full treatment against the average outcome under no treatment:
\begin{equation}\label{eq:tte}
\TTE = \frac{1}{N} \sum_{i=1}^{N} \left[ \E[Y_T^i(\mathbf{1})] - \E[Y_T^i(\mathbf{0})] \right],
\end{equation}
where $\mathbf{1}$ denotes the all-treated assignment and $\mathbf{0}$ denotes the all-control assignment across all time periods. This estimand captures both the direct effect of treatment on treated units and the indirect spillover effects that propagate through the network and over time.

In bipartite settings, treatment is applied to units on one side of the graph (e.g., service providers), while these treatment units interact with units on the other side, which we refer to as \emph{connected units} throughout this paper. We note that \citet{tan2025estimating}, and broadlly the bipartite-experiments literature, refer to these connected units as ``outcome units''; we deliberately avoid this terminology to prevent confusion, for reasons we now explain. In bipartite experimental designs, outcomes are often observed at the \emph{edge level}, that is, at each interaction between a treatment unit and a connected unit. For example, when a service provider serves a user, an outcome may be recorded for that specific interaction, such as the quantity of service delivered. These edge-level outcomes sometimes can then be aggregated to either side of the bipartite graph, referred by the \textit{linear additive edges} assumption in \citep{tan2025estimating}. One can aggregate to the treatment side by summing all edge outcomes across connected units served by each treatment unit, or aggregate to the connected-unit side by summing all edge outcomes across treatment units serving each connected unit. In this paper, we focus on outcomes aggregated to the treatment side: whenever we refer to outcomes $Y_t^i$, we mean the aggregate outcome for treatment unit $i$ at time $t$, computed by summing the edge-level outcomes across all connected units that treatment unit $i$ interacts with. This aggregation convention is why we avoid calling the right-hand-side units ``outcome units,'' as doing so would conflate the units with the outcomes themselves and obscure the aggregation structure. 

A further distinction comes up regarding which treatment units are eligible for the intervention. \citet{tan2025estimating} emphasize that treatment units often are of two type: those eligible for treatment and those ineligible due to operational or design constraints. One of their main contributions is formalizing this structure and introducing two corresponding estimands. The \emph{Primary Total Treatment Effect} (PTTE) measures the impact on eligible treatment units when they receive the intervention, while the \emph{Secondary Total Treatment Effect} (STTE) captures how outcomes of ineligible treatment units may be affected through spillovers, even though these units cannot themselves be treated. In this paper, we focus exclusively on eligible units: our indexing $i = 1, \ldots, N$ refers only to treatment units that are eligible for the intervention (but we allow existence of additional ineligible units). Consequently, the total treatment effect we study corresponds to what \citet{tan2025estimating} call the PTTE, and this is precisely the estimand they report for their real experimental analyses. We adopt the same focus. Studying how one might estimate the STTE using temporal methods such as causal message passing is an interesting direction, but it lies beyond the scope of this paper.

\subsection{Network-Aware Benchmark: Bipartite Graph Methodology}

\begin{figure}
\centering
\begin{tikzpicture}[scale=0.6]
    \tikzstyle{eligibleset} = [draw=blue!70!black, rectangle, rounded corners, minimum width=2.2cm, minimum height=3.5cm, line width=1.5pt]
    \tikzstyle{ineligibleset} = [draw=gray!60, rectangle, rounded corners, minimum width=2.2cm, minimum height=2.5cm, line width=1.5pt, dashed]
    \tikzstyle{connectedset} = [draw=green!60!black, rectangle, rounded corners, minimum width=2.2cm, minimum height=5cm, line width=1.5pt]
    \tikzstyle{unit} = [circle, fill, minimum size=4pt, inner sep=0pt]
    
    \node[eligibleset] (Eligible) at (0,2) {};
    \node[above=0.1cm of Eligible, text=blue!70!black, font=\small\bfseries] {Eligible};
    
    \node[unit, blue!70!black] (e1) at (-0.4,3.2) {};
    \node[unit, blue!70!black] (e2) at (0.4,2.6) {};
    \node[unit, blue!70!black] (e3) at (-0.3,1.8) {};
    \node[unit, blue!70!black] (e4) at (0.3,1.0) {};
    
    \node[ineligibleset] (Ineligible) at (0,-2.5) {};
    \node[below=0.1cm of Ineligible, text=gray!60, font=\small\bfseries] {Ineligible};
    
    \node[unit, gray!60] (i1) at (-0.3,-1.8) {};
    \node[unit, gray!60] (i2) at (0.3,-2.5) {};
    \node[unit, gray!60] (i3) at (0,-3.2) {};
    
    \node[connectedset] (Connected) at (8,0) {};
    \node[above=0.1cm of Connected, text=green!60!black, font=\small\bfseries] {Connected Units};
    
    \node[unit, green!60!black] (c1) at (7.6,1.8) {};
    \node[unit, green!60!black] (c2) at (8.4,1.2) {};
    \node[unit, green!60!black] (c3) at (7.7,0.4) {};
    \node[unit, green!60!black] (c4) at (8.3,-0.3) {};
    \node[unit, green!60!black] (c5) at (7.6,-1.1) {};
    \node[unit, green!60!black] (c6) at (8.4,-1.8) {};
    
    \draw[gray!50] (e1) -- (c1);
    \draw[gray!50] (e1) -- (c2);
    \draw[gray!50] (e2) -- (c2);
    \draw[gray!50] (e2) -- (c3);
    \draw[gray!50] (e3) -- (c3);
    \draw[gray!50] (e3) -- (c4);
    \draw[gray!50] (e4) -- (c4);
    \draw[gray!50] (e4) -- (c5);
    
    \draw[gray!50, dashed] (i1) -- (c4);
    \draw[gray!50, dashed] (i1) -- (c5);
    \draw[gray!50, dashed] (i2) -- (c5);
    \draw[gray!50, dashed] (i2) -- (c6);
    \draw[gray!50, dashed] (i3) -- (c6);
    
    \draw[decorate, decoration={brace, amplitude=8pt}, thick] (-2.2,-4.2) -- (-2.2,4.2);
    \node[font=\small, align=center, rotate=90] at (-3.5,0) {Treatment Units};
    
\end{tikzpicture}
\caption{Bipartite structure in experimental settings. Treatment units (left) are partitioned into eligible units, which can be assigned to treatment, and ineligible units, which remain in control throughout the experiment. Connected units (right) interact with treatment units through edges representing service relationships. Outcomes are aggregated at the treatment unit level by summing edge-level outcomes. Solid edges connect eligible treatment units to connected units; dashed edges connect ineligible treatment units. In this paper, our analysis focuses on eligible treatment units, indexed $i = 1, \ldots, N$.}
\label{fig:bipartite}
\end{figure}
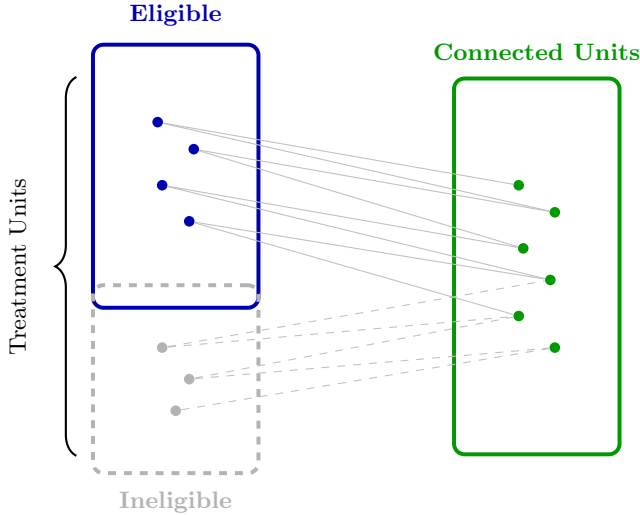

The network-aware approach of \citet{tan2025estimating} does not operate directly on temporal data. Instead, it works with outcomes aggregated across time at the unit level---specifically, the difference between post-treatment and pre-treatment average outcomes for each unit. This aggregated outcome structure is precisely what we refer to as the input to our SUTVA-based ``Basic Method.'' Treatment status is represented as a single indicator $W^{(j)} \in \{0,1\}$ for each treatment unit $j$, denoting whether that unit was assigned to treatment or control.

The methodology leverages the observed bipartite graph connecting treatment units and connected units. For each \emph{treatment unit} $j$, the method defines two exposure variables. The \emph{direct exposure} captures whether unit $j$ is itself treated and has connections in the graph:
\begin{equation}\label{eq:exposure-direct}
E^{\text{Dir}}_j(\mathbf{W}) = W^{(j)} \cdot |\mathcal{C}_j|,
\end{equation}
where $\mathcal{C}_j$ denotes the set of connected units linked to treatment unit $j$. The \emph{indirect exposure} captures spillover from other treated units through shared connected units:
\begin{equation}\label{eq:exposure-indirect}
E^{\text{Ind}}_j(\mathbf{W}) = \sum_{c \in \mathcal{C}_j} \sum_{k \neq j} W^{(k)} \cdot \mathbb{I}(k \in \mathcal{T}_c),
\end{equation}
where $\mathcal{T}_c$ denotes the set of treatment units connected to connected unit $c$, and $\mathbb{I}(\cdot)$ is the indicator function.

The outcome for treatment unit $j$ is modeled as a function of these exposures:
\begin{equation}\label{eq:outcome-treatment}
Y_j = \Psi(E^{\text{Dir}}_j, E^{\text{Ind}}_j, X_j) + \epsilon_j,
\end{equation}
where $X_j$ are covariates and $\Psi$ is estimated using machine learning methods such as kernel ridge regression. The PTTE is then estimated by comparing predicted outcomes under full treatment versus no treatment:
\begin{equation}\label{eq:ptte-network}
\widehat{\text{PTTE}} = \frac{1}{|\mathcal{T}|} \sum_{j \in \mathcal{T}} \left[ \hat{\Psi}(E^{\text{Dir}}_j(\mathbf{1}), E^{\text{Ind}}_j(\mathbf{1}), X_j) - \hat{\Psi}(0, 0, X_j) \right],
\end{equation}
where $\mathcal{T}$ denotes the set of eligible treatment units. A key requirement of this approach is observation of the bipartite graph structure linking treatment units and connected units.

\subsection{Causal Message Passing: Temporal Dynamics Without Network Data}

Causal message passing (CMP) takes a fundamentally different approach. Rather than conditioning on network structure, CMP exploits the \emph{temporal invariance} of interference mechanisms. The key insight is that treatment effects propagate through the network according to rules that remain structurally stable over time, analogous to invariant physical laws in interacting particle systems \citep{shirani2024causal, shirani2025can}.

CMP models the evolution of population-level outcome distributions through \emph{state evolution equations}:
\begin{equation}\label{eq:state-evolution}
Y_{t+1} \stackrel{d}{=} f_t(Y_t, W_{t+1}),
\end{equation}
where $Y_t$ and $W_{t+1}$ are single-dimensional random variables, representing the limiting distribution of outcomes and treatments at time $t$, respectively, while $f_t$ captures the aggregate dynamics of how outcomes evolve. The mapping $f_t$ incorporates both direct treatment effects and indirect spillover effects without requiring knowledge of individual network connections.

The estimation procedure proceeds in three steps:
\begin{enumerate}
    \item \emph{Feature construction:} Compute summary statistics of the outcome and treatment distributions at each time period (means, higher moments, interactions).
    
    \item \emph{State evolution estimation:} Use supervised learning to estimate the mappings $f_t$ from observed temporal variation in outcomes and treatments.
    
    \item \emph{Counterfactual prediction:} Apply the estimated mappings recursively to predict outcomes under alternative treatment scenarios (e.g., full treatment vs.\ no treatment).
\end{enumerate}

The total treatment effect is estimated by comparing the predicted counterfactual evolutions:
\begin{equation}\label{eq:tte-cmp}
\widehat{\TTE}^{\text{CMP}} = \widehat{\text{CFE}}_T(\mathbf{1}) - \widehat{\text{CFE}}_T(\mathbf{0}),
\end{equation}
where $\widehat{\text{CFE}}_T(\mathbf{w})$ denotes the estimated counterfactual evolution at the final time period under treatment allocation $\mathbf{w}$.

The key advantage of CMP is that it only requires observed outcomes over time, without any knowledge of the network data (i.e., $\omega_{ij}$'s used by the Network-Aware method). 

\subsection{Comparison of Approaches}

In the following section, we compare the Network-Aware and CMP methodologies against a simpler baseline that we call the \emph{Basic Method}. This Basic Method assumes SUTVA holds and ignores interference entirely. As discussed above, the Network-Aware approach operates on aggregated outcomes at the unit level, specifically the first differences between post-treatment and pre-treatment periods. 
The Basic Method uses the same aggregated outcome structure, collapsing the time series into a single pre-post difference for each unit, but applies a ML-enhanced version of standard difference-in-differences estimator that assumes SUTVA. By aggregating time before estimation, the Basic Method effectively discards the intermediate temporal dynamics that CMP leverages to disentangle interference.
The Network-Aware methodology can be viewed as a more sophisticated version of the Basic Method, one that incorporates the bipartite graph structure through exposure mappings. In contrast, CMP takes a fundamentally different approach by working directly with temporal outcome data and learning the dynamics of how treatment effects propagate over time.

Table~\ref{tab:comparison} summarizes the key differences among these three methodologies.

\begin{table}[h]
\centering
\caption{Comparison of Basic, Network-Aware, and CMP Approaches}
\label{tab:comparison}
\small
\begin{tabular}{lccc}
\toprule
\textbf{Feature} & \textbf{Basic} & \textbf{Network-Aware} & \textbf{CMP} \\
\midrule
Assumes SUTVA & Yes & No & No \\
Network data required & No & Yes & No \\
Temporal data required & No & No & Yes \\
Outcome structure & Aggregated & Aggregated & Time series \\
Exposure definition & None & Graph-based & Time-based \\
Estimation strategy & Diff-in-diff + ML & Propensity scores + ML & State evolution + ML \\
Identification source & Randomization only & Network structure & Temporal invariance \\
\bottomrule
\end{tabular}
\end{table}

\section{Experimental Setup}\label{sec:setup}

We apply all methodologies to two large-scale field experiments conducted in a service environment where treatment units interact with connected units through a bipartite structure. Due to confidentiality constraints by the data provider, we can only describe the experiments in general terms.

\subsection{Experimental Context}

Both experiments involve interventions applied to a subset of treatment-side units (e.g., drivers in a ride-sharing service). The bipartite structure arises naturally from the interaction patterns: connected units may connect to multiple treatment units, and treatment units serve multiple connected units.

Importantly, only a subset of treatment units are eligible for the intervention due to operational constraints. This creates an eligibility-constrained setting where randomization occurs within the eligible set, but ineligible units continue to interact with outcome units and may generate spillover effects.

The bipartite interaction graphs in both experimental settings are sparse, with the average degree of treatment units being small relative to the population size. This sparsity presents a non-trivial challenge for network-aware estimation, as the signal for exposure-response functions must be learned from relatively few interactions per unit.

\subsection{Experiments A and B}

Experiment A involved approximately 7,000 eligible treatment units and ran for multiple ($T<100$) time periods with varying treatment probabilities across experimental stages. The intervention modified a feature affecting how treatment units interact with connected units.

Experiment B involved approximately 4,000 eligible treatment units with a similar multi-stage design. The intervention targeted a related but different aspect of the interaction between the treatment and connected units.

Both experiments collected outcome data at the treatment unit level across all time periods, enabling application of both network-aware and CMP methodologies.

\subsection{Metrics}

We evaluate three pre-specified metrics, denoted $M_1$, $M_2$, and $M_3$:

\begin{itemize}
    \item \textbf{$M_1$:} A metric for which economic theory and domain knowledge predict that ignoring spillovers should bias estimates \emph{upward}, for both experiments A and B. This provides an ex-ante directional test.
    
    \item \textbf{$M_2$:} A secondary metric without a clear directional prediction for bias.
    
    \item \textbf{$M_3$:} The \emph{primary decision metric} used to evaluate the intervention's success. This metric determines whether the intervention would be launched at scale.
\end{itemize}

\subsection{Implementation}

We implement the bipartite methodology of \citet{tan2025estimating}, using the observed interaction graph to construct exposure variables and generalized propensity scores. LightGBM with parameter auto-tuning estimates the outcome function, with bootstrap inference for confidence intervals. We also implement the CMP framework of \citet{shirani2025can}, computing outcome and treatment summary statistics at each time period. We use ridge regression with cross-validation to estimate state evolution parameters, applying the distribution-preserving network bootstrap to generate training samples. We also implemented the Basic method described above that ignores interference.

\section{Results}\label{sec:results}

Table~\ref{tab:results} presents our main results, comparing the Basic (naive), Network-Aware, and CMP approaches across both experiments and all three metrics.

\begin{table}[htp]
\centering
\caption{Comparison of Basic Method, Network-Aware Approach, and Causal Message Passing on Two Real Experiments. For each metric, we report treatment effect estimates and statistical significance at the 5\% level. Bias columns show observed difference between Basic and Network-Aware, and expected bias direction from economic theory. $M_3$ is the primary decision metric.}
\label{tab:results}
\vspace{4mm}
\small
\begin{tabular}{@{}l@{\hskip 0.2cm}lcccccccc@{}}
\toprule
& \textbf{Metric} & \multicolumn{2}{c}{\textbf{Basic Method}} & \multicolumn{2}{c}{\textbf{Network-Aware}} & \multicolumn{2}{c}{\textbf{CMP}} & \multicolumn{2}{c}{\textbf{Bias}} \\
\cmidrule(lr){3-4} \cmidrule(lr){5-6} \cmidrule(lr){7-8} \cmidrule(lr){9-10}
& & ATE & Sig. & PTTE & Sig. & Effect & Sig. & Observed & Expected \\
\midrule
\textbf{Experiment A} & & & & & & & & & \\
\addlinespace[0.1em]
& $M_1$ & \textcolor{ForestGreen}{Pos.} & Yes & \textcolor{BrickRed}{Neg.} & No & \textcolor{BrickRed}{Neg.} & Yes & \textcolor{ForestGreen}{Pos.} & \textcolor{ForestGreen}{Pos.} \\
& $M_2$ & \textcolor{ForestGreen}{Pos.} & No & \textcolor{BrickRed}{Neg.} & No & \textcolor{BrickRed}{Neg.} & Yes & \textcolor{ForestGreen}{Pos.} & — \\
& $M_3$ & \textcolor{BrickRed}{Neg.} & No & \textcolor{ForestGreen}{Pos.} & Yes & \textcolor{ForestGreen}{Pos.} & Yes & \textcolor{BrickRed}{Neg.} & — \\
\addlinespace[0.2em]
\midrule
\addlinespace[0.1em]
\textbf{Experiment B} & & & & & & & & & \\
\addlinespace[0.1em]
& $M_1$ & \textcolor{ForestGreen}{Pos.} & No & \textcolor{ForestGreen}{Pos.} & No & \textcolor{ForestGreen}{Pos.} & Yes & \textcolor{ForestGreen}{Pos.} & \textcolor{ForestGreen}{Pos.} \\
& $M_2$ & \textcolor{ForestGreen}{Pos.} & Yes & \textcolor{ForestGreen}{Pos.} & No & \textcolor{ForestGreen}{Pos.} & No & \textcolor{ForestGreen}{Pos.} & — \\
& $M_3$ & \textcolor{ForestGreen}{Pos.} & No & \textcolor{BrickRed}{Neg.} & No & \textcolor{BrickRed}{Neg.} & No & \textcolor{ForestGreen}{Pos.} & — \\
\bottomrule
\end{tabular}
\end{table}

\subsection{CMP Recovers Network-Aware Estimates}

The most striking finding is the close alignment between CMP and Network-Aware estimates across both experiments and all metrics. In every case, CMP produces effect estimates with the \emph{same directional sign} as the Network-Aware approach. This alignment is notable given that CMP operates without any knowledge of the bipartite interaction structure that the Network-Aware method explicitly leverages.

For Experiment A, both Network-Aware and CMP find:
\begin{itemize}
    \item Negative effects on $M_1$ and $M_2$, reversing the postive estimates from the Basic method
    \item Positive and \emph{statistically significant} effects on the primary decision metric $M_3$
\end{itemize}

For Experiment B, both methods find:
\begin{itemize}
    \item Positive effects on $M_1$ and $M_2$
    \item Negative (non-significant) effects on $M_3$
\end{itemize}

CMP reaches statistical significance on metric $M_1$ where Network-Aware does not. One possible explanation is the difference in data utilization: while the Network-Aware method collapses the temporal dimension into a single pre-post difference per unit, CMP utilizes the full time series of outcomes. However, the two methods differ in many other respects, modeling assumptions, estimation procedures, and sources of identifying variation, making it difficult to attribute the difference in significance to any single factor. A rigorous comparison of statistical efficiency between these fundamentally different approaches is beyond the scope of this paper.

While CMP matches Network-Aware perfectly on the primary decision metric $M_3$, agreeing in both sign and statistical significance across both experiments, the two methods differ in statistical significance on metric $M_1$: CMP finds significant effects in both experiments where Network-Aware does not. Whether this reflects a genuine difference in statistical efficiency, differences in the estimands being targeted, or other methodological factors cannot be determined from our comparison alone. What we can conclude is that temporal dynamics provide sufficient signal to match the directional conclusions of the network-aware approach on all metrics studied.

\subsection{Correct Recovery of Bias Direction}

For metric $M_1$, economic theory predicts that ignoring spillovers should bias estimates in a specific direction: the Basic method should yield estimates that are more positive than the true total treatment effect. Both experiments confirm this prediction, and importantly, CMP's significant findings on this metric align with theoretical expectations.

In Experiment A, the Basic method shows a positive and significant effect, while both Network-Aware and CMP show negative effects. The difference (Basic $-$ Network-Aware) and (Basic $-$ CMP) are both positive, matching the expected bias direction. CMP's finding of a significant negative effect, where Network-Aware finds a non-significant negative effect, is consistent with domain knowledge about this metric. If the true effect is indeed negative as theory predicts, this pattern would suggest CMP may have greater power to detect it, though confirming this hypothesis would require further investigation.

In Experiment B, while all methods show positive effects, the observed bias (Basic more positive than the interference-aware estimates) still matches the expected direction. Again, CMP finds significance where Network-Aware does not. If the true effect is positive, this would be consistent with CMP having greater sensitivity, though other explanations cannot be ruled out.

This pattern demonstrates that CMP, like the Network-Aware approach, correctly identifies and adjusts for the bias that arises when spillovers are ignored. For metric $M_1$, where domain knowledge provides strong directional predictions, CMP's significant findings align with theoretical expectations. The temporal information that CMP extracts appears to be a reliable substitute for network information in these experimental settings, though understanding the precise trade-offs between temporal and network-based approaches remains an important direction for future research.

\subsection{Decision-Relevant Implications}

The primary decision metric $M_3$ reveals the practical importance of accounting for interference.

In Experiment A:, the Basic method suggests a negative treatment effect that does not reach statistical significance. In contrast, both Network-Aware and CMP find \emph{positive and statistically significant} effects. This reversal is consequential: a statistically significant increase in $M_3$ implies a deterioration on the primary decision metric, which the Basic method would fail to detect. The interference-aware methods provide a clear signal that the naive analysis obscures. Importantly, CMP and Network-Aware agree exactly on this decision-critical metric, same sign and significance conclusion.

In Experiment B, both Network-Aware and CMP find negative but non-significant effects on $M_3$, while the Basic method finds positive but non-significant effects. Although the directional disagreement exists, all methods lead to the same practical conclusion: no statistically significant effect detected. The interference-aware methods again agree with each other.

The agreement between CMP and Network-Aware on $M_3$ across both experiments is the key validation result of this paper. For the metric that determines whether an intervention should be launched at scale, the two methodologies, one requiring network data, one requiring only temporal data, reach identical conclusions.

\section{Discussion and Conclusion}\label{sec:discussion}

The strong alignment between CMP and Network-Aware estimates, particularly the perfect agreement on the primary decision metric $M_3$, supports the theoretical premise of causal message passing: that temporal variation in outcomes encodes information about interference structure. When treatment effects propagate through a network, they induce predictable patterns in how outcome distributions evolve over time. By learning these patterns, CMP effectively reconstructs the aggregate effect of network interference without observing the network itself.

On metric $M_1$, CMP reaches statistical significance where Network-Aware does not. While one plausible explanation is that CMP's use of outcome data across all time periods effectively increases the sample available for estimation, the two methods differ in too many dimensions to draw definitive conclusions about relative efficiency. Understanding the trade-off between network information and temporal information, and the conditions under which one dominates the other, is an important question that our empirical comparison raises but cannot resolve. What our results do suggest is that practitioners need not view the absence of network data as a fundamental obstacle to interference correction.

We hypothesize that the observed agreement arises when interference mechanisms are relatively stable over time, sufficient temporal variation exists to identify state evolution mappings, and the network structure generates interference patterns that manifest clearly in outcome dynamics. Future work should characterize conditions under which the methods may diverge and develop diagnostics for practitioners.

Several limitations warrant acknowledgment. Our validation is based on two experiments, and generalization to other domains requires additional study. Neither method provides the ``true'' treatment effect; we compare methods against each other and against directional predictions from theory. CMP requires multi-period outcome data and assumes relatively stable interference mechanisms, rapidly changing network structures may violate these assumptions. \cite{shirani2025evolution} provides an axiomatic treatment that can help understand such limitations better.

For practitioners facing settings where network data is unavailable, costly, or unreliable, CMP offers a principled alternative that correctly recovers bias direction and, in Experiment A, reverses a misleading conclusion that would arise from standard SUTVA-based analysis. More broadly, this work demonstrates empirical convergence between network-aware and network-blind methods on real data, suggesting that the information content of temporal dynamics and network structure may be more closely related than previously recognized.

\bibliographystyle{plainnat}
\bibliography{references}

\end{document}